\documentstyle[preprint,revtex]{aps}
\begin{document}
\draft
\preprint{}
\begin{title}
Effective Group Structure in the Interacting Boson Model
\end{title}
\author{N. D. Whelan}
\begin{center}
Centre for Chaos and Turbulence Studies

Niels Bohr Institute, Blegdamsvej 17, DK-2100, Copenhagen \/{O}, Denmark
\end{center}
\begin{abstract}
Spectrum generating algebras are used in various fields of physics as models
to determine quantum structure, including energy levels and transition
strengths.
The advantage of such models is that their group structure allows an extensive
understanding of the system being studied. In addition they
possess a simple classical limit, at least for bosonic systems. In this paper
we discuss an algebraic model of nuclear structure, the Interacting Boson
Model (IBM), and show that in one limit its group structure is particularly
simple.   For zero angular momentum there is an effective lower dimensional
group structure which describes the system both classically and quantum
mechanically.
\end{abstract}


\narrowtext
\section{Introduction}
\label{sec:intro}

Algebraic models have been developed to describe systems in several fields of
physics.  Examples in nuclear
physics include the Interacting Boson Model (IBM) \cite{ibm}
and its many extensions, the
vibron model \cite{vibron}, and the $SO(8)$ \cite{so8} and $Sp(6)$
\cite{so8,sp6} fermion dynamical symmetry models.  There exist similar
models in chemical physics \cite{chem1,chem2}
and hadronic physics \cite{part}.  Such models are typically
constructed by restricting the dynamics to a few important degrees of
freedom.  There is often an invariant group $G$ such that
any Hamiltonian we construct commutes with the group Casimir operators.
For the IBM this group is $U(6)$. One also often imposes the condition that
any Hamiltonian be invariant with respect to a
lower dimensional group $G'$ which is a subgroup of $G$. Often $G'$ is
$SO(3)$, implying spherical symmetry. For a given representation of $G$, the
Hamiltonian is block diagonal, with each block corresponding to a
different representation of $G'$.  In this paper, we show that it is possible
for one block to have a particularly simple structure.

We limit the discussion to the IBM.
We first find the classical limit of the
original quantum problem. This is a dynamical system with six degrees
of freedom.  Imposing zero angular momentum suppresses three of these degrees
of freedom \cite{paar,thesis},
leaving three.  The group structure of this reduced classical system is then
studied leading to the identification of
what we call an effective group structure. In particular, there
is an effective group $G_{\rm eff} = U(3)$ which is analogous to $G=U(6)$ of
the original system. We then quantise this reduced classical model to obtain
a new quantum model.  That this new model is different from the original is
not contradictory since more than one quantum problem can share the same
classical limit. Nevertheless, the solution of this new model
is similar to that of the original in the $J=0$ representation.
The idea that $J=0$ states correspond to a smaller dimension in
the classical mechanics was noticed in a very different context in
reference (\cite{scc}).

This result is of some interest in the study of chaos in collective
nuclei \cite{paar,thesis,ibmchaos,appr}. The three dimensional system is
further reduced by boson number conservation so that there are only
two independent degrees of freedom. Two dimensional systems are
particularly amenable to the usual analyses of chaos since the
classical motion can be depicted on Poincar\'{e} sections and
periodic orbits are easier to find. We can also study the quantum
mechanical wavefunctions and Husimi \cite{husi} distributions in
detail. It is then possible to
study wavefunctions and look for localisation on unstable periodic
orbits \cite{heller} among other effects.

In section 2 we briefly describe the IBM including its classical limit. The
group chains of the effective $U(3)$ model are discussed in section 3 in the
context of the classical dynamics.  In section 4 we discuss the allowed
representations of the effective group chains.  We impose a constraint that
our model only admit states which belong to the symmetric representation of
the permutation group $S_3$. It is then shown that the representations have
the same structure as in the original $U(6)$ model. In section 5 we discuss a
specific classical Hamiltonian and show how it can be integrated at the three
dynamical symmetries of the effective model.  In section 6 we describe a
quantum Hamiltonian with the required $U(3)$ symmetry and which corresponds
to the original $U(6)$ Hamiltonian.  We find its eigenenergies at the
dynamical symmetry limits and show that they are equal to
the energies of the original model to leading order in the particle number.
Therefore this can be thought of as a semiclassical approximation.
The philosophy of this approach is similar to that of reference \cite{hl}.

\section{The Interacting Boson Model}
\label{sec:ibm}

In this section we review those features of the IBM which are of special
interest in this paper. See reference [1] for a complete review.
The IBM is a model for the structure of even-even
collective nuclei which assumes that the monopole and quadrupole degrees of
freedom are the most important. It also assumes that all excitations are
bosonic because of the existence of pairing interactions which are dominant at
low energies.  Therefore, we introduce one monopole boson operator
$s^{\dagger}$ and five quadrupole boson operators $d^{\dagger}_{\mu}$
where $\mu=-2,\ldots,2$.

The 36 bilinear operators
$\{s^{\dagger}s,s^{\dagger}d_{\mu},d^{\dagger}_{\mu}s,d^{\dagger}_{\mu}d_{\nu}
\}$ are the generators of a $U(6)$ algebra.  This means that any Hamiltonian
constructed from the generators will commute with the $U(6)$ Casimir operators.
Since the system is bosonic, we only consider symmetric representations so
there is only one independent Casimir operator,
\begin{equation} \label{eq:qnum}
\hat{N}=s^{\dagger}s + \sum_{\mu}d^{\dagger}_{\mu}d_{\mu}
\end{equation}
The eigenvalue of this operator, $N$, is
the number of bosonic pairs or half the number of valence nucleons.
Any eigenstate of such a Hamiltonian belongs to a specific $U(6)$
representation which is labelled by $N$.

In addition to the $U(6)$ symmetry, we demand spherical symmetry by requiring
that there be $SO(3)$ invariance.
There are three group chains which connect these two groups
\begin{equation} \label{eq:chain}
U(6) \supset \left\{
\begin{array}{c}
U(5) \supset O(5)\\
SU(3)            \\
O(6) \supset O(5)
\end{array}
\right\} \supset SO(3) \supset SO(2).
\end{equation}
If we construct a Hamiltonian solely out of the Casimir invariants of one
chain, then it is solvable and we say that it has a dynamical symmetry.

The classical limit of the quantum model is obtained
\cite{hl,cllim2,cllim1,cllim3}
through the use of coherent states \cite{klaud}. These are defined as
\begin{equation} \label{eq:cohstate}
|\mbox{\boldmath $\alpha$}\rangle=
\exp(-|\mbox{\boldmath $\alpha$}|^{2}/2)\exp\Bigl(\alpha _{s}s^{\dagger}+
\sum_{\mu}\alpha_{\mu}d_{\mu}^{\dagger}\Bigr)|0\rangle
\end{equation}
where $|0\rangle$ is the vacuum state which contains no bosons.
Each coherent state is parametrised by six continuous complex variables
$\alpha_{j}$, where $j \in (s,-2,\ldots,2)$. Under time evolution a coherent
state $|\mbox{\boldmath $\alpha$}\rangle$ will evolve to a new state which is
approximately another coherent state  $|\mbox{\boldmath $\alpha'$}\rangle$.
Therefore, it is sufficient to study the time dependence of the variables
$\alpha_{j}$. A time-dependent variational
approximation valid for large boson number \cite{klaud,bo}
shows that these variables evolve according to Hamilton's equations
\begin{equation}
 \frac{d\alpha_j}{dt}=
\frac{\partial {\cal H} (\mbox{\boldmath $\alpha$})}
{\partial (i \alpha_j^*)} \;\;\;\;\;\;\;\;
\frac{d(i\alpha_j^*)}{dt}=-\frac{\partial {\cal H}
(\mbox{\boldmath $\alpha$})} {\partial \alpha_j}
\end{equation}
where
${\cal H} (\mbox{\boldmath $\alpha$}) =
\langle\mbox{\boldmath $\alpha$}|
\hat{H}|\mbox{\boldmath $\alpha$}\rangle$ is the classical
Hamiltonian and $\alpha_j$ and $i\alpha^*_j$ are canonical position and
momentum coordinates. In what follows we will denote quantum operators with
carats and their classical counterparts with script font.
The classical limit of any quantum operator, including the Hamiltonian, is
obtained by making the substitutions \cite{klaud,bo}
\begin{equation} \label{eq:dtoa}
d^{\dagger}_{\mu} \rightarrow  \alpha^{*}_{\mu} \;\;\;\;\;\;
d_{\mu}           \rightarrow  \alpha_{\mu}     \;\;\;\;\;\;
s^{\dagger}       \rightarrow  \alpha^{*}_{s}   \;\;\;\;\;\;
s                 \rightarrow  \alpha_{s}.
\end{equation}
For example, the classical limit of the number operator (\ref{eq:qnum}) is
${\cal N} = \alpha_s^* \alpha_s + \sum_{\mu}\alpha_\mu^*\alpha_\mu$. The fact
that it is conserved means that the phase space is compact.

This prescription of finding the classical limit is not unique. For example,
the coherent states defined in equation (\ref{eq:cohstate}) do not belong to
a specific $U(6)$
or $SO(3)$ representation.  We might like them to have this property so that
in the classical dynamics we are studying a reduced dimensional phase space
on which the classical boson number and angular momentum are fixed. This has
been done for the $U(6)$ algebra \cite{cllim1} in which use of projected
coherent states
means that there is one fewer degree of freedom in the classical dynamics.
It has not been done for the $SO(3)$ algebra; presumably the resulting
phase space would be very topologically complicated.  In this work we will
show that we can define a simple phase space for the special case of
zero angular momentum.
Another method of finding the classical limit is to identify the quantum
commutators with the classical Poisson brackets \cite{moyal}.
One then identifies each of
the generators as a phase space coordinate and uses the Casimir invariants to
eliminate variables; this has been done explicitly for the $SU(2)$ and $SU(3)$
algebras \cite{dim}. This is an elegant approach but is difficult in this
application because the bosonic nature of the IBM only allows
symmetric representations of $U(6)$ and it is not clear how to invoke
this constraint using this method.

In reference \cite{hl} the authors discuss how the variables
$\alpha_j$ can be related to the intrinsic variables of the
Bohr-Mottelson model \cite{bohrmot}.
These are the deformation parameters $\beta$ and
$\gamma$, the Euler angles $\Omega$ and the corresponding momenta $p_{\beta}$,
$p_{\gamma}$ and ${\cal L}_{1,2,3}$. The result is
\begin{mathletters} \label{eq:alpha}
\begin{eqnarray}
\alpha_s & = & \exp (-i\Theta)\sqrt{{\cal N}-\frac{1}{2}
\sum_{\mu}(p^*_{\mu}p_{\mu} + q^*_{\mu}q_{\mu})} \label{equationa}\\
\alpha_{\mu} & = & \exp (-i\Theta)(q^*_{\mu}+ip_{\mu}) \nonumber
\end{eqnarray}
where
\begin{eqnarray}
q_{\mu} & = & \sum_{\nu=-2}^2{\cal D}_{\mu\nu}^{(2)}(\Omega)a_{\nu}
\label{equationb}\\
p_{\mu} & = & \sum_{\nu=-2}^2{\cal D}_{\mu\nu}^{*(2)}(\Omega)b_{\nu}
\nonumber
\end{eqnarray}
and
\begin{equation} \label{equationc}
a_{\pm 2} = \frac{1}{\sqrt{2}}\beta\sin\gamma
\;\;\;\;\;\;\;\;\;\;
a_{\pm 1} = 0
\;\;\;\;\;\;\;\;\;\;
a_{0} = \beta\cos\gamma
\end{equation}
\begin{eqnarray}
b_{\pm 2} & = & \frac{1}{\sqrt{2}}\Bigl(\frac{p_{\gamma}}{\beta}\cos\gamma +
p_{\beta}\sin\gamma \pm \frac{i{\cal L}_3}{2\beta\sin\gamma}\Bigr) \nonumber\\
b_{\pm 1} & = & -\frac{1}{2\sqrt{2}\beta}\Bigl(\frac{i{\cal L}_1}
{\sin (\gamma-2\pi/3)} \pm \frac{{\cal L}_2}{\sin(\gamma-4\pi/3)}\Bigr)
\label{equationd}\\
b_0 & = & -\frac{p_{\gamma}}{\beta}\sin\gamma + p_{\beta}\cos\gamma.\nonumber
\end{eqnarray}
\end{mathletters}
Here ${\cal D}_{\mu\nu}^{(2)}(\Omega)$ are the Wigner matrices which are a
function of the three Euler angles $\Omega$ and $\Theta$ is a global phase
of no importance. This choice of variables explicitly conserves ${\cal N}$.

There is a simplification when the magnitude of the angular momentum
is zero \cite{paar,thesis}.
In that case we have that the ${\cal L}_i$ are zero and we can choose a frame
in which ${\cal D}_{\mu\nu}^{(2)}(\Omega)= \delta_{\mu\nu}$. Equation (6)
then implies
{\samepage
\begin{eqnarray}
\alpha_s & = & \exp (-i\Theta)\sqrt{{\cal N}-\frac{1}{2}(\beta^2 + p_{\beta}^2
+ \frac{p_{\gamma}^2}{\beta^2})}    \nonumber  \\
\alpha_{\pm 2} & = & \frac{1}{2}\Bigl(\beta\sin\gamma +
i(\frac{p_{\gamma}}{\beta}\cos\gamma + p_{\beta}\sin\gamma)\Bigr)
\label{eq:alphas} \\
\alpha_{\pm 1} & = & 0 \nonumber  \\
\alpha_0 & = & \frac{1}{\sqrt{2}}\Bigl(\beta\cos\gamma +
i(-\frac{p_{\gamma}}{\beta}\sin\gamma + p_{\beta}\cos\gamma)\Bigr).\nonumber
\end{eqnarray}
}
\noindent Then the motion is described by only two degrees of freedom, $\beta$
and $\gamma$.
For example, $\hat{H} = \hat{n}_d$ is a quantum Hamiltonian belonging to
the $U(5)$ dynamical symmetry. ($\hat{n}_d$ is the $U(5)$ Casimir operator and
equals $\sum d^{\dagger}_{\mu}d_{\mu}$.) Its classical limit for zero angular
momentum is ${\cal H} = (\beta^2+p_{\beta}^2+p_{\gamma}^2/\beta^2)/2$ which is
a harmonic oscillator in two dimensions. We have ignored terms
which arise from normal ordering since they vanish in the classical limit.
Therefore, the original $U(5)$ symmetry is manifest as a $U(2)$ symmetry in
this situation. We show in the next section that all of
the groups in the original model map to lower dimensional groups when we study
the classical problem with zero angular momentum.

This dimensional reduction can also be understood on the quantum level by
counting the number of distinct eigenvalues which is necessary to specify a
quantum state. For a fixed value of $N$ we need five quantum numbers to do
this.
We always have angular momentum $J$ and one of its components $M$ as good
eigenvalues.  Therefore, in general we need three additional quantum numbers so
that the system is three dimensional \cite{cllim3}. However, in the special
case
that $J=0$, it follows that $K$, the projection of $J$ onto the symmetry axis,
is also zero. ($K$ is also a
``missing quantum number'' in the decomposition from $SU(3)$ to $SO(3)$.)
Therefore, once we specify $J=0$, we
only need to specify two additional quantum numbers to identify a state,
which means that the system is essentially two dimensional.

\section{Effective Group Chains}
\label{sec:efg}

In this section, we discuss the various group chains
which arise in classifying
the classical behaviour for zero angular momentum. First, it is convenient to
define new variables
\begin{equation} \label{eq:alphapm}
\alpha_{p,m} = \frac{1}{\sqrt{2}}(\alpha_2\pm\alpha_{-2}).
\end{equation}
Equation (\ref{eq:alphas}) then guarantees that $\alpha_m=\alpha_{\pm 1}=0$.
After making this canonical transformation we are free to use the fact that
$\alpha_m=0$ so that $\alpha_{\pm}=\alpha_p/\sqrt{2}$.
For now, we will explicitly
keep $\alpha_s$ and not use ${\cal N}$ conservation to eliminate it.
We then have a problem in three degrees of freedom given by $\alpha_s$,
$\alpha_0$ and $\alpha_p$ and can define nine bilinear objects
\begin{equation}
g_{ij} = \alpha^*_i\alpha_j \;\;\;\;\;\; i,j\in (s,0,p).
\end{equation}
Their Poisson brackets are
$\{g_{ij},g_{kl}\}= \frac{1}{i}(g_{il}\delta_{jk} - g_{kj}\delta_{il})$, which
is the classical version \cite{moyal} of the quantum commutator relation
$[\hat{g}_{ij},\hat{g}_{kl}]= (\hat{g}_{il}\delta_{jk} -
\hat{g}_{kj}\delta_{il}).$ These are the
commutator relations for the generators of the $U(3)$ algebra and
we conclude that classically we have a $U(3)$ algebra.  The Casimir invariant
$\sum_i\alpha_i^*\alpha_i$, has zero Poisson bracket with any Hamiltonian we
construct from the generators and is therefore always a
constant of motion. This is
just the particle number and is numerically equal to ${\cal N}$.

We wish to consider the nature of the possible Hamiltonians which can be
constructed from
these nine generators. These Hamiltonians are integrable if they have three
independent constants in involution \cite{licht}.
This is guaranteed if there is a dynamical symmetry \cite{cllim3,feng},
so we want to know the possible dynamical symmetries.  To find these,
it is helpful to refer to the original $U(6)$ model. Its first group chain is
obtained by considering a $U(5)$ subalgebra which consists of 25 generators.
Of these generators, all but four are zero for zero angular momentum.
The remaining ones are
\begin{equation}
\begin{array}{ccc}
\alpha^*_p\alpha_p & \;\;\;\;\;\;\;\;\;\; & \alpha^*_0\alpha_p\\
\alpha^*_p\alpha_0 & \;\;\;\;\;\;\;\;\;\; & \alpha^*_0\alpha_0
\end{array}
\end{equation}
which have a $U(2)$ algebra.  We can consider an $SO(2)$ subalgebra whose
generator is $g_3=i(\alpha^*_p\alpha_0 - \alpha^*_0\alpha_p)$. The reason for
this choice will be discussed below. Therefore, we can identify one
group chain as
\begin{equation} \label{eq:u2}
U(3) \supset U(2) \supset SO(2)
\end{equation}
If we write down a Hamiltonian in terms of the Casimir invariants of these
groups, then these invariants are constants of motion in involution and the
motion is integrable.

The second group chain of the $U(6)$ model is $SU(3)$ which has eight
generators \cite{ibm}.
The first three are the three components of the angular momentum
and the other five are components of the rank two quadrupole tensor,
\begin{equation} \label{eq:qclass}
{\cal Q}_{\mu} = \alpha_s^*\tilde{\alpha}_{\mu} +
\alpha_{\mu}^*\alpha_s
- \frac{\sqrt{7}}{2}(\alpha^*\tilde{\alpha})^{(2)}_{\mu}
\end{equation}
where the last term means that we couple the objects in the brackets to
angular momentum two and $\tilde{\alpha}_{\mu}=(-1)^{\mu}\alpha_{-\mu}$.
In the present case all of the angular momenta components are zero.  In
addition, ${\cal Q}_{\pm 1} = 0$ and ${\cal Q}_2={\cal Q}_{-2}$ so we are
left with two independent quantities
\begin{eqnarray}
q_0 & = &  \alpha^*_s\alpha_0 + \alpha^*_0\alpha_s -
\frac{1}{\sqrt{2}}(\alpha^*_p\alpha_p - \alpha^*_0\alpha_0)\\
q_2 & = &  \alpha^*_s\alpha_p + \alpha^*_p\alpha_s -
\frac{1}{\sqrt{2}}(\alpha^*_p\alpha_0 + \alpha^*_0\alpha_p),\nonumber
\end{eqnarray}
where $q_0={\cal Q}_0$ and $q_2=\sqrt{2}{\cal Q}_{\pm 2}$.
It is straightforward to show that $q_0$ and $q_2 $ have zero Poisson bracket
so they are the generators of a $U(1) \times U(1)$ algebra. We can therefore
identify the group chain as
\begin{equation} \label{eq:u1u1}
U(3) \supset U(1) \times
U(1).
\end{equation}

The third group chain is $O(6)$ in the original model.  It has 15
generators. There are three components of the angular momentum, which are
zero.  There are five components of a quadrupole tensor which is the same as
in equation (\ref{eq:qclass}) but without the last term. As in the
$SU(3)$ case ${\cal Q}_{\pm 1}=0$ and ${\cal Q}_2={\cal Q}_{-2}$, so only two
components are independent.  Finally, there are seven components of a rank
three octupole
tensor ${\cal O}$. These are all zero except the $\mu=\pm 2$ components which
are equal.  Therefore, we have three independent quantities
\begin{eqnarray} \label{eq:so3gen}
g_1 & = & \alpha^*_0\alpha_s + \alpha^*_s\alpha_0  \nonumber\\
g_2 & = & \alpha^*_p\alpha_s + \alpha^*_s\alpha_p \\
g_3 & = & i(\alpha^*_p\alpha_0 - \alpha^*_0\alpha_p)\nonumber
\end{eqnarray}
with $g_1={\cal Q}_0$, $g_2=\sqrt{2}{\cal Q}_{\pm 2}$ and $
g_3=2i{\cal O}_{\pm 2}$. These have an $SO(3)$ structure.
We can consider $g_3$ to be the generator of an $SO(2)$ algebra, as in the
$U(2)$ chain.  We then have the group chain
\begin{equation} \label{eq:o3}
U(3) \supset SO(3) \supset SO(2).
\end{equation}

In summary, the original $U(6)$ group chain structure as shown in equation
(\ref{eq:chain}) has the following simpler structure when we consider the
classical model for zero angular momentum:
\begin{equation} \label{eq:chaineff}
U(3) \supset \left\{
\begin{array}{c}
U(2) \supset SO(2)\\
U(1) \times U(1)\\
SO(3) \supset SO(2).
\end{array}
\right.
\end{equation}

A few comments are in order.  In the original $U(6)$ model, there is a
constraint that all group chains must contain the $SO(3)$ subalgebra for
which
the $\alpha_{\mu}^*$ are rank two spherical tensors.  For example, this
constrains the $U(5)$ subalgebra to be that one which contains all the
$\alpha_{\mu}^{*}\alpha_{\nu}$ generators and none of the generators which
contain $\alpha_s$ or $\alpha_s^*$. An analogous constraint on the allowed
representations in this picture is that all states must be in a symmetric
representation of the $S_3$ permutation group, as will be discussed.
Another important point is that for the classical mechanics, these group
chains are exact. The Poisson brackets of the relevant degrees of freedom
have precisely the structure appropriate for the three chains derived above.
There are slight problems upon requantisation, as will be discussed below.

It is also useful to motivate the choice of $SO(2)$ generator for the $U(2)$
and $SO(3)$ algebra chains.  Consider the original $O(5)$ Casimir invariant
\cite{ibm}
\begin{equation}
{\cal C}_2(O(5)) = 4\Bigl(
(\alpha^*\tilde{\alpha})^{(3)}\cdot(\alpha^*\tilde{\alpha})^{(3)}
+ (\alpha^*\tilde{\alpha})^{(1)}\cdot(\alpha^*\tilde{\alpha})^{(1)}
\Bigr).
\end{equation}
The second term is zero for zero angular momentum. Also
$(\alpha^*\tilde{\alpha})^{(3)}_{\pm 1,\pm 3} = 0$ since no combination of
$\alpha_0$ or $\alpha_{\pm 2}$ can combine to give an odd index.  In addition
$(\alpha^*\tilde{\alpha})^{(3)}_0 = 0$ due to the equality of $\alpha_2$ and
$\alpha_{-2}$.  This leaves
\begin{equation}
(\alpha^*\tilde{\alpha})^{(3)}_{\pm 2} = \pm\frac{1}{2}
(\alpha^*_p\alpha_0 - \alpha^*_0\alpha_p).
\end{equation}
Therefore,
\begin{eqnarray}
{\cal C}_2(O5) & = & -2(\alpha^*_p\alpha_0 - \alpha^*_0\alpha_p)^2 \\
               & = &  2g_3^2, \nonumber
\end{eqnarray}
where $g_3$ is precisely the $SO(2)$ generator in the first and third group
chains. In fact, equations (\ref{eq:alphas}) and (\ref{eq:so3gen}) imply
$g_3=p_{\gamma}$ so that
$g_3$ generates $\gamma$ rotations.  Thus, $SO(2)$ symmetry is the same as
$\gamma$ independence and implies $p_{\gamma}$ conservation. In the
zero angular momentum limit,
this result agrees with the well know connection between the $O(5)$
Casimir invariant and $p_{\gamma}$ \cite{hl}.
It should be emphasised that the $SO(3)$ and $SO(2)$ groups discussed here
are not groups of spatial rotations but are groups describing abstract
transformations among the bosons.

\section{Representations of the Effective Groups}

In this section, we discuss the allowed representations of the three
group chains.  We start with the first group chain which is given by equation
(\ref{eq:u2}). $U(3)$ is labelled by the quantum number $N$; this is the same
as the $U(6)$ eigenvalue since in each case $N$ refers to the number of bosons.
The $U(2)$ representation label is denoted
by $n$ and can take all integer values from
0 to $N$.  For a given representation of $U(2)$, the allowed representations
of $SO(2)$ are $\mu = -n, -n+2,\ldots,n.$
This selection of every other representation is a general property of the group
chain $U(d)\supset SO(d)$. For $d=2$, it is derived in reference
\cite{messiah} in a discussion of two dimensional harmonic oscillators.

However, not all of these $SO(2)$ representations are physically realised.
There is an additional constraint that all wavefunctions are in a symmetric
representation of $C_{3v}\approx S_3$, the three point permutation group
\cite{bohrmot}. This is most easily seen in the configuration
space shown in Fig.~1 for which $\beta$ and
$\gamma$ are polar coordinates. Since $U(3)$ is
a compact group there is a constraint that $\beta\leq 2$. The heavy lines
at $\gamma=0$, $2\pi/3$ and $4\pi/3$ denote prolate symmetry and the dashed
lines at $\gamma=\pi/3$, $\pi$ and $5\pi/3$ denote oblate symmetry. There are
six physically indistinguishable regions connected by the six elements of
the $C_{3v}$ group. This means that all states must belong to the symmetric
representation of this group. We can think of the configuration space as
being given by just one of the six domains. This is not the case for general
angular momentum, since the angular momentum axis picks out a direction which
differentiates among the domains.

More fundamentally, the fact that the system has the symmetry of the three
point permutation group arises from the fact that for zero angular
momentum no axis is special so any relabelling of the $x$, $y$ and $z$
axes leaves the system invariant.  Such relabellings correspond to spatial
rotations and reflections. The fact that the zero angular momentum
states belong to the symmetric representation of the $S_3$ group arises because
they must be invariant with respect to all spatial rotations and reflections
and therefore with respect to all relabellings.  Only the symmetric
representation has this property so it follows that the zero angular
momentum states belong to the symmetric representation of $S_3$.  Therefore,
the condition of having zero angular momentum specifies both the $S_3$
symmetry and the relevant representation.

It is a simple result of group theory \cite{gt} that the functions
$\cos 3\mu\gamma$, with $\mu$ a non-negative integer, belong to the symmetric
$A$ representation of $C_{3v}$. (The functions $\sin 3\mu\gamma$ belong to the
$B$ representation and all other possibilities belong to the two dimensional
$E$ representation.) Thus, we can label the allowed representations of $SO(2)$
by $\mu$ non-negative and divisible by three. This prescription yields
precisely the same number of allowed states as the original $U(6)$ problem.
This is simple enough to show in general but is most easily demonstrated
with an example.

Consider the $U(3)$ representation $N=6$.  The possible representations of
the first group chain are shown in Table~I. For comparison, we also show
the representation labels for the first group chain
of the original $U(6)$ model \cite{ibm} in the $J=0$ representation as shown
in equation (\ref{eq:chain}). The $U(5)$ algebra is labelled by the eigenvalue
$n_d$, the $O(5)$ algebra is labelled by $v$ and the missing quantum number
in going from $O(5)$ to $O(3)$ is identically zero and is not considered.
In each case, the number of allowed states is the same.  What is more, the
representation labels are identical. This is a general result; the
representation labels of the effective groups can be identified with
the labels of the original model.

Since the third group chain is similar to the first, we discuss it next. Recall
that it is given by equation (\ref{eq:o3}). Following the earlier discussion,
the allowed $SO(3)$ representations include every other
integer counting down from $N$; we label them by the quantum number $I$.
The decomposition from $SO(3)$ to $SO(2)$ follows the normal rule so we have,
\begin{eqnarray}
I   & = & N,N-2,\ldots,\mbox{1 or 0} \\
\mu & = & -I,-I+1,\ldots,I\nonumber
\end{eqnarray}
However, as above, we only consider $\mu$ non-negative and divisible
by three.  The results for $N=6$ are shown in Table~II.  We have seven states
as in the previous group chain.  The multiplicities of
the $SO(2)$ representations are the same for both group chains; in
each case we have $\mu=0^4,3^2,6$. The left half of the table shows the result
for the $J=0$ representation of the third group chain shown in equation
(\ref{eq:chain}).  $O(6)$ and $O(5)$ representations are labelled by
$\sigma$ and $v$ respectively.  We again find that the group labels are the
same.

We conclude by considering the second group chain. We are
constrained to select only those representations which are symmetric in $S_3$.
To do this it is convenient to think of this group chain as
\begin{equation}
\begin{array}{ccccc}
U(3) & \hspace{.3cm}   & \supset & \hspace{.3cm} & U(1)\times U(1)\times
U(1).\\
  N  &                 &         &               &
n_s \hspace{1cm} n_+ \hspace{1cm} n_-
\end{array}
\end{equation}
where each number labels the representation of the group above it.
Then our states are $|n_s\rangle |n_+\rangle |n_-\rangle$ with $N=n_s+n_++n_-$.
In this case, the
$S_3$ group acts to interchange the labels so that the representation which is
completely symmetric under this group is
\begin{eqnarray}
|\phi(n_s,n_-,n_+)\rangle & = &\frac{1}{\sqrt{6}}\Bigl(
|n_s\rangle |n_+\rangle |n_-\rangle +
|n_-\rangle |n_s\rangle |n_+\rangle +
|n_+\rangle |n_-\rangle |n_s\rangle \\
& & + |n_s\rangle |n_-\rangle |n_+\rangle
    + |n_+\rangle |n_s\rangle |n_-\rangle
    + |n_-\rangle |n_+\rangle |n_s\rangle\Bigr).\nonumber
\end{eqnarray}

We can label the states with three integers
\begin{equation} \label{eq:nval}
k \geq l \geq m \;\;\;\;\;\;\;\;\;  k+l+m=N
\end{equation}
such that the states are the symmetric combination of distributing the integers
$(k,l,m)$ among the eigenvalues $(n_s,n_-,n_+)$. Our representations are then
\begin{eqnarray}
m & = & 0,1,\ldots,\left[\frac{N}{3}\right] \nonumber\\
l & = & m,m+1,\ldots,\left[\frac{N-m}{2}\right] \label{eq:ineq}\\
k & = & N - l - m\nonumber
\end{eqnarray}
where $[x]$ is the smallest integer less than or equal to $x$.
This guarantees that the
conditions (\ref{eq:nval}) are satisfied. It is convenient to define two
other labels
\begin{eqnarray}
a & = & 2(k-l)    \label{eq:pq}\\
b & = & 2(l-m).   \nonumber
\end{eqnarray}
By equation (\ref{eq:ineq}) these are both non-negative.

We can now determine the allowed representations for this group chain with
$N=6$. The seven pairs of values of $a$ and $b$ are shown in Table~III. Also
shown are the representations of the second
group chain of equation (\ref{eq:chain}) which are
labelled by two eigenvalues $(\lambda,\mu)$.  As before, the representation
labels are identical. In terms of Young tableaux, the procedure
for selecting the allowed representations of $U(1)\times U(1)$ is
identical to that of finding the allowed representations of $SU(3)$
\cite{gt}.

In conclusion, we see that within the $U(3)$ picture there is a very
clear way to find the allowed representations of each group chain.  These
labels are identical to those of the original $U(6)$ model.  At this level,
the correspondence is exact since the effective group
chains give the same numbers of states with the same multiplicities
and the same group labels.
However, once
we try to calculate quantum eigenvalues, we will see that there are
differences and this $U(3)$ picture will be shown to be a leading order
semiclassical approximation.

\section{The Classical Hamiltonian}

    It is helpful at this stage to describe a classical Hamiltonian which can
interpolate among the three dynamical symmetries.  One choice is the extended
consistent-Q Hamiltonian \cite{cw88} whose classical limit for arbitrary
angular momentum is \cite{thesis,appr}
\begin{equation}
{\cal H} = \eta{\cal NU} - (1-\eta){\cal Q}^{\chi}\cdot{\cal Q}^{\chi}.
\end{equation}
This depends on two parameters, $\eta$ and $\chi$. The quantity ${\cal U}$ is
the classical analogue of $\hat{n}_d$.  For zero angular momentum,
we have ${\cal Q}_{\pm 1}=0$ and ${\cal Q}_2={\cal Q}_{-2}$ so the Hamiltonian
reduces to
\begin{equation} \label{eq:cham}
{\cal H} = \eta{\cal NU} -
(1-\eta)\Bigl({\cal Q}_0^2(\bar{\chi}) + 2{\cal Q}_2^2(\bar{\chi})\Bigr)
\end{equation}
with
\begin{eqnarray}
{\cal U} & = & \alpha_0^*\alpha_0 + \alpha_p^*\alpha_p \nonumber\\
{\cal Q}_0(\bar{\chi}) & = & \alpha_0^*\alpha_s + \alpha_s^*\alpha_0 -
\frac{\bar{\chi}}{\sqrt{2}}(\alpha_p^*\alpha_p - \alpha_0^*\alpha_0)
\label{eq:cterms}\\
{\cal Q}_2(\bar{\chi}) & = & \Bigl(\alpha_p^*\alpha_s + \alpha_s^*\alpha_p -
\frac{\bar{\chi}}{\sqrt{2}}(\alpha_p^*\alpha_0 + \alpha_0^*\alpha_p)\Bigr)
/\sqrt{2}. \nonumber
\end{eqnarray}
where we have made use of the substitution (\ref{eq:alphapm}).
The first term in equation (\ref{eq:cterms}) describes vibrations and the
second describes
quadrupole interactions.  The parameter $\bar{\chi}$ is related to the more
commonly used $\chi$ by the relation $\bar{\chi} = \chi/(-\sqrt{7}/2)$.
$n_d$ is a 1-body term which scales as $\cal N$ while the quadrupole term is
2-body and scales as ${\cal N}^2$. Therefore we have multiplied the first term
by $\cal N$ \cite{thesis,appr} so that both terms scale the same.
This Hamiltonian can also be expressed in terms of the intrinsic coordinates
defined in equation (\ref{eq:alphas}) for which we find \cite{paar,thesis,hl}
${\cal U}=(p_{\beta}^2
+\beta^2+p_{\gamma}^2/\beta^2)/2$ and
{\samepage
\begin{eqnarray}
{\cal H} & = & \eta{\cal N}{\cal U} -(1-\eta)\Bigl[2\beta^2({\cal
                 N}-{\cal U}) \nonumber\\
         &   & -\bar{\chi}\sqrt{{\cal N}-{\cal U}}
                 \Bigl((p_{\gamma}^2/\beta-\beta p_{\beta}^2-\beta^3)\cos
                 3\gamma + 2p_{\beta}p_{\gamma}\sin 3\gamma\Bigr) \\
         &   &  +\frac{\bar{\chi}^2}{2}({\cal U}^2 - p_{\gamma}^2)\Bigr].
\nonumber
\end{eqnarray}}

The Hamiltonian for $\bar{\chi}=0$ has the reasonably simple form
\begin{equation}
{\cal H} = \eta{\cal N}{\cal U} -2(1-\eta)\beta^2({\cal N}-{\cal U}).
\end{equation}
This is clearly $\gamma$-independent so it has an $SO(2)$ symmetry for all
values
of $\eta$.  This is enough to ensure integrability because ${\cal N}$,
${\cal H}$ and $p_{\gamma}^2$ are three constants of motion in involution.
The Hamiltonian has additional $U(2)$ and $SO(3)$ symmetries at
the values $\eta=1$
and $\eta=0$ respectively.  These limits are called over-integrable.
Two familiar examples of over-integrable systems in three dimensions are
the harmonic oscillator and the Coulomb potential. These have $U(3)$
and $SO(4)$ symmetries respectively \cite{golds} even though the $SO(3)$
symmetry of spherical rotations is enough to ensure integrability.

Another case of interest is $\bar{\chi}=1$ and $\eta=0$, for which
the Hamiltonian has a $U(1)\times U(1)$ symmetry.  This is not manifest if we
consider the Hamiltonian in the original coordinates $\alpha_j$, but a change
of variables makes it clear \cite{thesis}.  Define new coordinates $\zeta_j$
by
\begin{eqnarray}
\alpha_s & = & (\zeta_s-\zeta_+ -\zeta_-)/\sqrt{3} \nonumber\\
\alpha_0 & = & (2\zeta_s + \zeta_+ +\zeta_-)/\sqrt{6} \\
\alpha_p & = & (\zeta_+ - \zeta_-)/\sqrt{2}.\nonumber
\end{eqnarray}
This transformation is both unitary and canonical. $\zeta_s=
(\alpha_s+\sqrt{2}\alpha_0)/\sqrt{3}$ is the classical analog of the $SU(3)$
boson condensate \cite{ami}.

Straightforward algebra leads to
\begin{equation} \label{eq:hu1u1}
{\cal H} = -2({\cal J}_s^2+{\cal J}_+^2+{\cal J}_-^2-{\cal J}_+{\cal J}_-
-{\cal J}_+{\cal J}_s-{\cal J}_-{\cal J}_s),
\end{equation}
where ${\cal J}_i=\zeta_i^*\zeta_i$.
It is clear that $\{{\cal J}_i,{\cal H}\}=0$, so
the three ${\cal J}_i$ constitute a set of independent constants of motion.
Their
existence implies integrability; in fact they are precisely the action
coordinates of the problem \cite{thesis}.
We can use the fact that ${\cal N} = {\cal J}_s+{\cal J}_++{\cal J}_-$
to obtain
\begin{equation}
{\cal H} =
-\Bigl[2{\cal N}^2 -6{\cal N}({\cal J}_++{\cal J}_-)
+ 6({\cal J}_+^2+{\cal J}_-^2+{\cal J}_+{\cal J}_-)\Bigr].
\end{equation}

It is now clear that the meaning of the $U(1)\times U(1)$ symmetry is that
there are two independent oscillations
\begin{equation}
\zeta_{\pm}(t) = \zeta_{\pm}(0)\exp(i\Omega_{\pm}t)
\end{equation}
with frequencies
\begin{equation}
\Omega_{\pm} = 6{\cal N} - 12{\cal J}_{\pm} - 6{\cal J}_{\mp}.
\end{equation}
The frequencies depend explicitly on the amplitudes of the motion. For small
oscillations thy are approximately degenerate with values
close to $6{\cal N}$. This corresponds to vibrations near the potential
minimum at $\gamma=0$ and $\beta=\sqrt{4{\cal N}/3}$
which are approximately degenerate \cite{hl}.

\section{The Quantum Hamiltonian}

We next consider the effect of quantising the classical system discussed
above. As a first step we will define boson creation and annihilation
operators by reversing equation (\ref{eq:dtoa})
\begin{equation} \label{eq:atod}
\alpha_p^*       \rightarrow     d^{\dagger}_p    \;\;\;\;\;
\alpha_p         \rightarrow     d_p              \;\;\;\;\;
\alpha_0^*       \rightarrow     d^{\dagger}_0    \;\;\;\;\;
\alpha_0         \rightarrow     d_0              \;\;\;\;\;
\alpha^*_s       \rightarrow     s^{\dagger}      \;\;\;\;\;
\alpha_s         \rightarrow     s.
\end{equation}
In general, there are ordering ambiguities in going from a classical to a
quantum Hamiltonian.  Here a natural ordering suggests itself. We started with
a $U(6)$ quantum Hamiltonian, found its classical limit and then recognised
that
there was an effective $U(3)$ chain which describes zero angular momentum.
At no point was it
necessary to change the ordering of terms in the Hamiltonian.  Therefore, we
will take the $U(3)$ quantum Hamiltonian with the same ordering as the
original quantum Hamiltonian.

The result of quantising equations (\ref{eq:cham}) and (\ref{eq:cterms}) is
\begin{mathletters} \label{eq:qham}
\begin{equation} \label{eqa}
\hat{H} = N\eta\hat{n}_d - (1-\eta)\Bigl(\hat{Q}^2_0(\bar{\chi})
+2\hat{Q}^2_2(\bar{\chi})\Bigr)
\end{equation}
where
\begin{eqnarray}
\hat{n}_d & = & d^{\dagger}_0 d_0 + d^{\dagger}_p d_p\nonumber\\
\hat{Q}_0(\bar{\chi}) & = & d^{\dagger}_0s + s^{\dagger}d_0 -
    \frac{\bar{\chi}}{\sqrt{2}}(d^{\dagger}_pd_p - d^{\dagger}_0d_0)
\label{eqb}\\
\hat{Q}_2(\bar{\chi}) & = & \Bigl(d^{\dagger}_ps + s^{\dagger}d_p -
    \frac{\bar{\chi}}{\sqrt{2}}(d^{\dagger}_pd_0 + d^{\dagger}_0d_p)\Bigr)
/\sqrt{2} \nonumber
\end{eqnarray}
\end{mathletters}
It is worth stressing that this quantum Hamiltonian was obtained in a three
step
process.  We first found the classical limit of the $U(6)$ model.  We then
used a dimensional reduction appropriate for zero angular momentum.  Finally,
we requantised using a natural ordering.  The effect of these three steps is
formally the same as substituting
\begin{equation}
\begin{array}{ccc}
d_{\pm 1}  = 0 & \;\;\;\;\;\;\;\;\;\;\;\;\;\;\; & d^{\dagger}_{\pm 1} = 0 \\
d_2 = d_{-2}  &  \;\;\;\;\;\;\;\;\;\;\;\;\;\;\; &
d^{\dagger}_2 = d^{\dagger}_{-2}
\end{array}
\end{equation}
in the original Hamiltonian.
We know that these substitutions are inconsistent
because Heisenberg's uncertainty principle implies that two noncommuting
variables can not be simultaneously equal to zero.  The error is in the
requantisation since we should include the effects of the zero point motion
of the three neglected degrees of freedom.  These contribute in the next to
leading order term in $\hbar$ (since $\hbar\sim 1/N$ \cite{bo}).
The conclusion is that
we have defined a new quantum problem of lower dimension whose solution is
a semi-classical approximation to the original $U(6)$ problem. In reference
\cite{hl} the authors found a similar ${\cal O}(1/N)$ discrepancy after a
similar analysis.

We next discuss the eigenenergies of the $U(3)$ Hamiltonian above.
For $\eta=1$, we have $\hat{H} = N\hat{n}_d$ which is proportional to the
linear $U(2)$ Casimir operator.  The energies are then
\begin{equation}
E_i = Nn_{di}
\end{equation}
where $i$ indicates the i'th state and $n_{di}$ labels its $U(2)$
representation.  This result agrees exactly with the result for the original
$U(6)$ vibrational limit. This is not a general result but arises because
there are no ordering ambiguities in our choice of Hamiltonian. Consider the
more common choice for a vibrational
classical Hamiltonian in $d$ dimensions; $H=\sum_i(p_i^2+x_i^2)/2$.
Its quantum energies are $E=\sum_i n_i+d/2$.  In this case, neglecting
degrees of freedom would result in a discrepancy in the second term. We will
see that such higher order discrepancies are the rule, not the exception.
As mentioned, the semiclassical limit is obtained for large
$N$.  All group labels are of order $N$, so the eigenvalues of the quantum
Hamiltonian ((\ref{eq:qham})
have leading order terms which are quadratic in the
group labels and the next order terms are linear.

We next consider the third group chain which corresponds to
$\eta=\bar{\chi}=0$. In that case,
\begin{eqnarray}
\hat{H} & = & -\Bigl(\hat{Q}_0^2(0)+2\hat{Q}_2^2(0)\Bigr)\\
        & = & -(\hat{g}_1^2 + \hat{g}_2^2) \nonumber
\end{eqnarray}
where the quantum generators $\hat{g}_1$, $\hat{g}_2$ and $\hat{g}_3$ are
obtained by applying the substitution (\ref{eq:atod}) to the classical
generators given in equation (\ref{eq:so3gen}). We then have
\begin{equation}
\hat{H} = -\hat{C}(O3) + \hat{C}(O2) \nonumber
\end{equation}
so that
\begin{equation} \label{eq:o3en}
E_{I\mu}      =  -I(I+1) + \mu^2.
\end{equation}
The result for the original model is
\begin{equation} \label{eq:o6en}
E_{\sigma v} = -\sigma(\sigma+4) + v(v+3),
\end{equation}
where $\sigma$ and $v$ label the $O(6)$ and $O(5)$ representations,
respectively.
Recalling the identification of $I$ and $\mu$ with $\sigma$ and $v$, we
see that this is a leading order approximation to the energies.

It is possible to make an argument about the next order term for the
energies of this dynamical symmetry.
Consideration of the radial Schr\"{o}dinger equation in $d$ dimensions gives
the semi-classical approximation for the eigenvalues of the $SO(d)$ Casimir
operator
\cite{jim}
\begin{equation} \label{eq:sod}
C\Bigl(SO(d)\Bigr) = \left(I+\frac{d-2}{2}\right)^2
\end{equation}
where $I$ is the integer which labels the $SO(d)$ representation. The special
case of three dimensions gives the well known Langer modification \cite{lang}
\begin{eqnarray}
C\Bigl(SO(3)\Bigr) & \approx & \left(J+\frac{1}{2}\right)^2 \\
        & = & J(J+1) + \frac{1}{4}.\nonumber
\end{eqnarray}
This disagrees with the exact result by a factor of $1/4$, which is a third
order term.  The $(d-2)/2$ factor in equation (\ref{eq:sod}) can be thought of
as arising from turning points and their phase space generalisations
\cite{gutz}. We can then interpret equation (\ref{eq:o3en}) as
\begin{equation} \label{eq:o3appr}
E_{I\mu} \approx -\Bigl(I+\frac{d-2}{2}\Bigr)^2 +
\Bigl(\mu+\frac{(d-1)-2}{2}\Bigr)^2
\end{equation}
with $d=3$.  This agrees with equation (\ref{eq:o3en}) in the first two terms
and the dimension $d$ only enters into the second leading term.  To account
for the suppressed degrees of freedom, it is reasonable to add 3 to the
dimension.  Substituting $d=6$ into formula (\ref{eq:o3appr}) gives
\begin{eqnarray}
E_{I\mu} & = & -(I+2)^2 + (\mu+\frac{3}{2})^2 \\
         & = & -I(I+4) + \mu(\mu+3) - \frac{7}{4}. \nonumber
\end{eqnarray}
As promised, this agrees with the original result (\ref{eq:o6en})
for the leading two terms.
In practise, this is sufficient since constants are unimportant when
calculating the differences between energies.  However, it should be stressed
that this is only a plausibility argument and is not rigorous.

We next discuss the energy eigenvalues of the second dynamical symmetry.
Its classical Hamiltonian, in the coordinates $\zeta$, is given by equation
(\ref{eq:hu1u1}). We quantise by replacing these variables by creation and
annihilation operators.  The Hamiltonian is then a function  of the group
Casimir operators.  These are $U(1)$ operators whose eigenvalues are
integers.  We then obtain the energies
\begin{equation}
E_{n_s n_+ n_-} = -2(n_s^2+n_+^2+n_-^2 - n_s n_+ - n_s n_- - n_+ n_-).
\end{equation}
We can also obtain this by a direct semiclassical approximation of equation
(\ref{eq:hu1u1}) by substituting ${\cal J}_i=n_i$ which is appropriate for
complex phase space \cite{voros}. It is interesting to note that in this
situation the semiclassical approximation is exact.  However there are
still  semiclassical errors arising from the dimensional reduction as
shown in the next paragraph.

Expressing this in terms of $a$ and $b$ defined in equation (\ref{eq:pq})
leads to
\begin{equation}
E_{ab} = -\frac{1}{2}(a^2 + b^2 + ab).
\end{equation}
The exact result for the $SU(3)$ limit of the original model is
\begin{equation}
E_{\lambda\mu} = -\frac{1}{2}(\lambda^2 + \mu^2 +\lambda\mu +3\lambda+3\mu).
\end{equation}
where $\lambda$ and $\mu$ label the $SU(3)$ representation.  Earlier it was
argued that $a$ and $b$ can be identified with $\lambda$ and $\mu$ so we see
that the previous two formulae agree to leading order.

An important point is that the approximation for all the group chains is
valid if $N$ is large but with no added constraint.  Therefore, this
procedure reproduces, approximately, the entire spectrum for a given value
of $N$.  This is in contrast to the results of reference \cite{hl}
where the approximate energies are valid for $N$ large and for the other
quantum numbers much smaller than $N$.  This reproduces the energies of
the low-lying states but not of the entire spectrum.

In principle, it should be possible to derive the next order terms to the
energies by more sophisticated semiclassical arguments. However, for our
purposes it is not important since the leading order terms are
sufficient to identify the quantum states with the effective group
representations. It is this correspondence which is the central result of
this paper.

\section{Conclusion}

The point which is established in this paper is that
if a Hamiltonian has a group structure and is block diagonal, then
at least one of the blocks may have a simpler effective group structure.
This was motivated by studying the classical mechanics for which the
effective group chains are unambiguous.  The quantum
mechanics is a little more troublesome since it is inconsistent to
completely ignore degrees of
freedom; their zero point motions can be significant in understanding the
quantum structure. Nevertheless, we have shown that it is possible to
requantise within the picture of
the effective group structure.  Arguments about the required point
group symmetry of the states limit the allowed representations so that
the number of states is the same as in the original model. Furthermore,
we can identify the representation labels of the effective groups with the
labels of the original groups.  Therefore, at this level even the quantum
mechanics of the effective groups is unambiguous; it is completely consistent
to identify quantum states
with representations of the lower dimensional effective
groups. Only at the point of calculating quantum energies is there a problem.
Because we have not consistently accounted for the missing degrees of freedom,
there are higher order corrections to the energies derived here. The
interesting problem of how to get the higher order terms remains.

This is a useful result because it means that we have a simple understanding
of the $J=0$ quantum states. They can be defined in a two
dimensional space and their dynamical symmetries have a clear, intuitive
interpretation.  The case of zero angular momentum has also been of interest
in studying chaos \cite{paar,thesis,appr,ibmchaos} since the two dimensional
nature of the classical motion makes the analysis of the classical
dynamics simpler.  It is therefore useful to understand
the nature of the possible symmetries in the reduced problem.
It is also possible to calculate quantum energies away from the dynamical
symmetry limits within this effective group picture.  For example, it
is quite simple to diagonalise a general Hamiltonian in a basis of
$U(2)$ eigenstates. This has been done and the results will be discussed in a
subsequent publication.

There are two interesting features of the IBM model which will probably be
easier to understand for zero angular momentum. The first is that there is a
suppression of chaos for a family of Hamiltonians between the $SU(3)$ and
$U(5)$ limits \cite{thesis,appr}. We can now try to explain it as a
property of the
effective model between the $U(1)\times U(1)$ and $U(2)$ dynamical symmetries.
The other feature is that the IBM has a partial
dynamical symmetry \cite{prtl}.  This term describes a situation in which the
Hamiltonian does not have any symmetry and yet a subset of its quantum
eigenstates do.  This affects the classical dynamics by reducing the extent of
chaos \cite{alw}.  A semiclassical understanding
of this effect might be possible for zero angular momentum for which the
partial $SU(3)$ symmetry becomes a simpler partial $U(1)\times U(1)$ symmetry.
More generally, this may serve as a useful starting point to study
the effects of classical chaos on collective nuclear structure.

An interesting point is that the $U(3)$ model discussed here is
equivalent to the three level Lipkin model \cite{lipkin} in the case where the
number of fermions equals the number of single particle states in each level.
It is amusing to
note that this model, which provides a phenomenological tool for the
study of shell effects, might actually have a physical realisation in
collective nuclei.  However, unlike the usual studies involving the
Lipkin model, we have the additional constraint of being in a symmetric
$S_3$ representation. The Lipkin model has been studied in the context of
chaos \cite{meredith} but without explicit reference to its
dynamical symmetries.

This idea might also apply in other systems.  For example, there exist
models of triatomic molecules \cite{chem2} based on assuming a $U(4)
\times U(4)$ algebra and insisting on an $SO(3)$ subalgebra. There are various
dynamical symmetries, some of which involve an $SO(4)$ algebra.  Insisting
on the $J=0$ representation of $SO(3)$ limits the allowed $SO(4)$
representations.  Then the quantum states can be specified in terms of one
fewer quantum number so there is a dimensional reduction of one.  It might be
that this situation is also described by an effective group chain, but this
must
be worked out in detail. It would also be interesting to see if this idea can
be applied to fermionic systems for which it is difficult to refer to
the classical limit.

\acknowledgments

I would like to thank Yoram Alhassid and Ami Leviatan for useful discussions
and Jim Morehead for allowing me to use his unpublished calculations.
I also thank Peter Dimon for critical comments on the manuscript.
This work was supported under the EU Human Capital and Mobility Programme.

\figure{IBM configuration space.  The heavy lines denote prolate symmetry and
the dashed lines denote oblate symmetry.  The six segments are physically
identical; we arbitrarily choose the hatched region to be the configuration
space.  $\beta$ and $\gamma$ act as polar coordinates labelling the arbitrary
point $P$.}

\newpage
\mediumtext
\begin{table} \label{tab1}
\caption{Allowed representations of the first group chain of the $U(3)$ model
and of the $J=0$ $U(6)$ model. In both cases $N=6$. Each column refers to
one group and the letter in brackets is the group label used in the text.}
\begin{tabular}{cccc}
\multicolumn{2}{c}{$U(3)$} & \multicolumn{2}{c}{$U(6)$ with $J=0$}\\
\hline
$U(2)$ ($n$) & $SO(2)$ ($\mu$) & $U(5)$ ($n_d$) & $O(5)$ ($v$)\\
\hline
6 & 0,6 & 6 & 0,6\\
5 & 3   & 5 & 3  \\
4 & 0   & 4 & 0  \\
3 & 3   & 3 & 3  \\
2 & 0   & 2 & 0  \\
1 & -   & 1 & -  \\
0 & 0   & 0 & 0  \\
\end{tabular}
\end{table}

\begin{table}
\caption{The same as in Table~I but for the third group chain.}
\begin{tabular}{cccc}
\multicolumn{2}{c}{$U(3)$} & \multicolumn{2}{c}{$U(6)$ with $J=0$}\\
\hline
$SO(3)$ ($I$) & $SO(2)$ ($\mu$) & $O(6)$ ($\sigma$) & $O(5)$ ($v$)\\
\hline
6 & 0,3,6 & 6 & 0,3,6\\
4 & 0,3   & 4 & 0,3  \\
2 & 0     & 2 & 0    \\
0 & 0     & 0 & 0    \\
\end{tabular}
\end{table}

\begin{table}
\caption{The same as in Table~I but for the second group chain.}
\begin{tabular}{cc}
\multicolumn{1}{c}{$U(3)$} & \multicolumn{1}{c}{$U(6)$ with $J=0$}\\
\hline
$U(1)\times U(1)$ $(a,b)$ & $SU(3)$ $(\lambda,\mu)$ \\
\hline
$(12,0) \; (8,2) \; (4,4) \; (0,6)$ & $(12,0) \; (8,2) \; (4,4) \; (0,6)$\\
$(6,0) \; (2,2)                   $ & $(6,0) \; (2,2)                   $\\
$(0,0)                            $ & $(0,0)                            $\\
\end{tabular}
\end{table}

\end{document}